\documentclass[twocolumn,pra,aps,superscriptaddress,nofootinbib]{revtex4-1}
\usepackage{graphicx,bm,bbm,amsfonts,amsmath,amsthm,mathrsfs,amssymb}
\usepackage[usenames,dvipsnames]{color}
\usepackage{hyperref}
\usepackage{amsthm}

\newcommand{\cD}{\mathcal{D}}
\newcommand{\cE}{\mathcal{E}}
\newcommand{\cF}{\mathcal{F}}
\newcommand{\cG}{\mathcal{G}}

\newcommand{\cI}{\mathcal{I}}

\newcommand{\tr}{\mathrm{tr}}

\newcommand{\upi}{\mathrm{i}}

\newcommand{\id}{\mathbbm{1}}
\newcommand{\Fmin}{F_{\min}}
\newcommand{\Gnoisy}{\widetilde\cG}

\usepackage[normalem]{ulem}

\newcommand{\SetGrey}[1]{\definecolor{Grey}{gray}{#1}}
\SetGrey{0.60}

\begin{document}
\title{Direct estimation of minimum gate fidelity}
\author{Yiping Lu}
\affiliation{Yale-NUS College, Singapore 138527, Singapore}
\affiliation{Centre for Quantum Technologies, National University of Singapore, Singapore 117543, Singapore}
\author{Jun Yan Sim}
\affiliation{Centre for Quantum Technologies, National University of Singapore, Singapore 117543, Singapore}
\author{Jun Suzuki}
\affiliation{Graduate School of Informatics and Engineering, The University of Electro-Communications, Tokyo 182-8585, Japan}
\author{Berthold-Georg Englert}
\affiliation{Centre for Quantum Technologies, National University of Singapore, Singapore 117543, Singapore}
\affiliation{Department of Physics, National University of Singapore, Singapore 117551, Singapore}
\affiliation{MajuLab, CNRS-UCA-SU-NUS-NTU International Joint Research Unit, Singapore}
\author{Hui Khoon Ng}
\email{huikhoon.ng@yale-nus.edu.sg}
\affiliation{Yale-NUS College, Singapore 138527, Singapore}
\affiliation{Centre for Quantum Technologies, National University of Singapore, Singapore 117543, Singapore}
\affiliation{MajuLab, CNRS-UCA-SU-NUS-NTU International Joint Research Unit, Singapore}
\date{\today}

\begin{abstract}
With the current interest in building quantum computers, there is a strong need for accurate and efficient characterization of the noise in quantum gate implementations. A key measure of the performance of a quantum gate is the minimum gate fidelity, i.e., the fidelity of the gate, minimized over all input states. Conventionally, the minimum fidelity is estimated by first accurately reconstructing the full gate process matrix using the experimental procedure of quantum process tomography (QPT). Then, a numerical minimization is carried out to find the minimum fidelity. QPT is, however, well known to be costly, and it might appear that we can do better, if the goal is only to estimate one single number. In this work, we propose a hybrid numerical-experimental scheme that employs a numerical gradient-free minimization (GFM) and an experimental target-fidelity estimation procedure to directly estimate the minimum fidelity without reconstructing the process matrix. We compare this to an alternative scheme, referred to as QPT fidelity estimation, that does use QPT, but directly employs the minimum gate fidelity as the termination criterion. Both approaches can thus be considered as direct estimation schemes. General resource estimates suggest a significant resource savings for the GFM scheme over QPT fidelity estimation; numerical simulations for specific classes of noise, however, show that both schemes have similar performance, reminding us of the need for caution when using general bounds for specific examples. The GFM scheme, however, presents potential for future improvements in resource cost, with the development of even more efficient GFM algorithms.
\end{abstract}
\maketitle

\section{Introduction}
Quantum computing is much in the news these days, with the recent achievement of quantum supremacy by the Google device \cite{Naturegoogle} and the renewed interest in building a quantum computer (see, for example, Refs.~\cite{Nature1,Nature2}). The main obstacle to realizing a quantum computer of a useful scale and accuracy is the noise that threatens to destroy the quantum features that give quantum computers their power. A key focus of any implementation of a quantum information processing device is hence the characterization, and subsequent control and mitigation, of the noise that unavoidably accompanies the operation of a quantum gate.

An often used measure of the quality of a quantum gate implementation is the fidelity of the gate, i.e., a quantification of how close the action of the actual gate is to the theoretical ideal. Efficient and easy-to-implement procedures such as randomized benchmarking \cite{RB1,RB2,RB3} can offer information on the \emph{average} fidelity of a gate, averaged over input states according to some distribution of interest (e.g., Haar-distributed pure states for randomized benchmarking). What is often more telling about the gate performance, however, is the \emph{minimum} fidelity, i.e., the fidelity of the gate operation, minimized over all possible input states. This is a state-distribution-independent quantity, and gives a minimum guarantee for the quality of the gate, which is important to assure that the quantum computer functions correctly in all scenarios.

Conventionally, the minimum fidelity is estimated by first a full quantum process tomography (QPT) of the noisy gate operation, followed by a numerical minimization, using the estimated process matrix, to find the minimum fidelity over all possible input states. This is a potentially costly procedure: For a $d$-dimensional system, QPT requires an accurate estimation of $d^4$ real parameters, all to yield one number, i.e., the minimum fidelity. Resource estimates of the number of uses of the channel required for an accurate full QPT can be deduced from bounds for state tomography \cite{Kueng2014,SampleComplexity}, giving $O(d^6/\epsilon'^2)$, where $\epsilon'$ is the accuracy of the reconstruction, measured by a distance between quantum processes.

One expects to be able to do better if the desire is only to estimate the minimum fidelity to some desired accuracy $\epsilon$, without the full reconstruction of the process matrix of the noisy gate. In this work, we propose a direct route to estimating the minimum fidelity, without the use of QPT. It combines a numerical gradient-free minimization (GFM) algorithm with the direct target fidelity estimation scheme of Refs.~\cite{DF1,DF2} to perform a hybrid numerical-experimental descent of the fidelity function to the minimum. We compare this GFM scheme with an alternative route, which we name \emph{QPT fidelity estimation}. QPT fidelity estimation does go through QPT, but rather than employing a process distance as the figure of merit as done in standard QPT, we make direct use of the minimum fidelity as the stopping rule. In this sense, this second route can also be regarded as a direct estimation of the minimum fidelity and offers fair comparison with the GFM scheme.

Resource estimates suggest that QPT fidelity estimation needs $O(d^8/\epsilon^2)$ uses of the gate to achieve an estimate of the minimum fidelity to accuracy $\epsilon$. This follows from an argument (see Sec.~\ref{sec:theory}) that relates the desired minimum fidelity accuracy of $\epsilon$ to the Choi-state trace-distance accuracy of $\epsilon'$ used in Refs.~\cite{Kueng2014,SampleComplexity}. This $O(d^8)$ uses of the gate, for fixed $\epsilon$, is a prohibitively high resource cost. The GFM approach, in contrast, is expected (see Sec.~\ref{sec:theory}) to require much milder $O(d^4)$ gate uses. It is thus, from this perspective, a much more efficient procedure for estimating the minimum fidelity.

For specific classes of noisy gates, however, our numerical results suggest a somewhat different conclusion. We numerically simulated both GFM and QPT fidelity estimation procedures for two natural classes of noisy gates: noisy gates random in the Hilbert--Schmidt sense, and gates with random Pauli and amplitude-damping noise. For both classes, our GFM scheme performs close to (though better than) the resource estimates; the QPT fidelity estimation scheme, surprisingly, took only $O(d^4)$ gate uses, much fewer than the $O(d^8)$ prediction. The performance of the two schemes is hence comparable for these two classes of noisy gates, with only slightly better scaling for the GFM scheme. This better performance for QPT is especially unexpected as our simulations do not follow the optimal procedure behind the theoretical bounds obtained in Refs.~\cite{Kueng2014,SampleComplexity}. This reminds us that, for specific classes of noisy gates, the resource estimates, which account even for worst-case scenarios, may not provide a reliable gauge of typical performance.

\makeatletter\global\advance\@colroom-10pt\relax\set@vsize\makeatother

Nevertheless, while there is little room for improvement in the resource scaling for QPT fidelity estimation, the scaling for the GFM scheme is limited largely by the efficiency of the numerical GFM algorithm, a subject of intense study in the field of numerical optimization. Our work also emphasizes the importance of making direct use of the quantity of interest, in this case the minimum fidelity, in the measurement procedure, rather than a secondary quantity such as a process distance. Both schemes explored here, the GFM scheme as well as the QPT fidelity estimation, can be considered when looking for methods of direct estimation of the minimum gate fidelity, with the GFM offering potential for further reduction in resource cost, while the QPT fidelity estimation offers the advantage of a measurement setup familiar from standard QPT.

Whether the observation of Ref.~\cite{Huang+2:2020}, namely, that very
  few measurements, chosen at random, may be sufficient for the estimation of
  particular properties, has a bearing on estimating the minimum gate fidelity
  is currently unknown, and this deserves to be explored. 
  The matter is, however, not within the scope of this work.

Below, we begin in Sec.~\ref{sec:directEst} with a description of our GFM algorithm, also providing a reminder of the direct target fidelity estimation scheme following the analysis of Ref.~\cite{DF1}. We then explain the QPT fidelity estimation procedure in Sec.~\ref{sec:QPT}. Section \ref{sec:perf} gives the resource estimates and numerical simulations that compare the performance of the two schemes. We conclude in Sec.~\ref{sec:conc}.

\section{Direct estimation with gradient-free minimization}\label{sec:directEst}
For a known ideal gate $\cG$, acting on an $n$-qubit system, and its noisy implementation $\Gnoisy$, we are interested in how close $\Gnoisy$ is to $\cG$. One way to quantify this is to compare the fidelity between the output state of $\Gnoisy$ with the ideal output state after $\cG$. The minimum fidelity, minimized over all input states, is defined as
\begin{equation}\label{eq:Fmin}
\Fmin\equiv\min_{\psi}F\bigl(\cG(\psi),\Gnoisy(\psi)\bigr)=\min_{\psi}F\bigl(\psi,\cE(\psi)\bigr).
\end{equation}
Here, $\psi\equiv |\psi\rangle\langle\psi|$ is a pure state on the $d(=\!2^n)$-dimensional Hilbert space of the quantum system, $F(|\phi\rangle\langle\phi|,\rho)\equiv \langle\phi|\rho|\phi\rangle$ denotes the squared fidelity between the pure state $|\phi\rangle$ and the (possibly mixed) state $\rho$, and $\cE\equiv \cG^{-1} \circ\Gnoisy$ is the noise process that describes the imperfections in the gate implementation $\Gnoisy$. $\cG^{-1}$ is the inverse of the ideal unitary gate $\cG(\cdot)=U(\cdot)U^\dagger$, for unitary $U$, i.e., $\cG^{-1}(\cdot)=U^\dagger(\cdot) U$. We assume that $\Gnoisy$ is a completely positive (CP) and trace-preserving (TP) map---also referred to as a quantum channel---and hence so is $\cE$. Even though the minimization in Eq.~\eqref{eq:Fmin} appears to only be over pure states $\psi$, $\Fmin$ is, in fact, the minimum fidelity over all states, pure or mixed, as the concavity of $F$ ensures that the minimum is attained on a pure state.

The goal here is to estimate $\Fmin$ without first estimating the full process matrix of the noisy gate $\Gnoisy$, or, equivalently, of the noise channel $\cE$. We assume the following experimental capabilities: (i) We can prepare any input state of our choice; (ii) we can send that input state through the noisy gate and access the output state; (iii) we can perform product-Pauli measurements on the output state. The noisy gate is regarded here as a black box $\Gnoisy$ in the laboratory that takes the input $\psi$ and gives back the output $\Gnoisy(\psi)$.

We need two additional ingredients. The first is the technique of direct estimation of target fidelity (DTFE) invented in Ref.~\cite{DF1}, allowing the estimation of the fidelity of an $n$-qubit state $\rho$ with some target pure state $|\psi\rangle$ without full state tomography. One needs only make probabilistic product-Pauli measurements, according to a distribution determined by the target pure state (see Sec.~\ref{sec:DTFE} for more details).

The second ingredient is a numerical method for gradient-free minimization (GFM), implementable on a classical computer. A GFM method finds a local minimum of a function in the case where function values, but not the gradient values, are easily available as inputs to the algorithm. A common situation is one where the function itself cannot be written down explicitly, but is accessible only through a numerical procedure. The gradient of the function is hence also not available as a function that can be written down explicitly, and methods of numerically approximating the gradient typically do not work well or are prohibitively expensive to evaluate.
GFM methods incorporate the gradient estimation with the minimization, by choosing trial points in the domain space, deducing some gradient information (often only a rough estimate) from the function values evaluated at those points, taking a small step in a direction expected to lower the function value according to that gradient information, and then repeating the process with new trial points at the new location. With standard regularity criteria on the function, one eventually arrives at a local minimum, and a repetition of the procedure sufficiently many times gives a good chance of finding the global minimum.

We now put the ingredients together for our GFM scheme of direct estimation of the minimum fidelity. Again, the goal is to estimate $\Fmin$ without full knowledge of $\Gnoisy$, apart from access to it as an input-output black box in the laboratory. As is typical in such problems, the requirement is to estimate $\Fmin$ to within a target accuracy of $\epsilon$ from the true value with high probability, i.e., $|\widehat\Fmin-\Fmin|\leq \epsilon$ for some small $\epsilon>0$, where $\widehat\Fmin$ is the estimate while $\Fmin$ is the (unknown) true value. The function we are minimizing here is $f(\psi)\equiv F\bigl(\psi,\cE(\psi)\bigr)$, over the domain of pure states $\psi$. A full description of $f$ and its gradient for minimization using gradient-descent methods requires knowledge of $\cE$, which we do not possess. However, we can evaluate the function value in the laboratory by preparing $\psi$, feeding it into the black box $\Gnoisy$, and then estimating $f(\psi)$ by carrying out the DTFE scheme in the laboratory, treating $\cG(\psi)$ as the target (pure) state and $\Gnoisy(\psi)$ as the state to be compared with the target. In this way, we have access, from measurements in the laboratory on the black box $\Gnoisy$, to the function values $f(\psi)$.

\begin{figure}[t]
\centering
\includegraphics[width=\columnwidth]{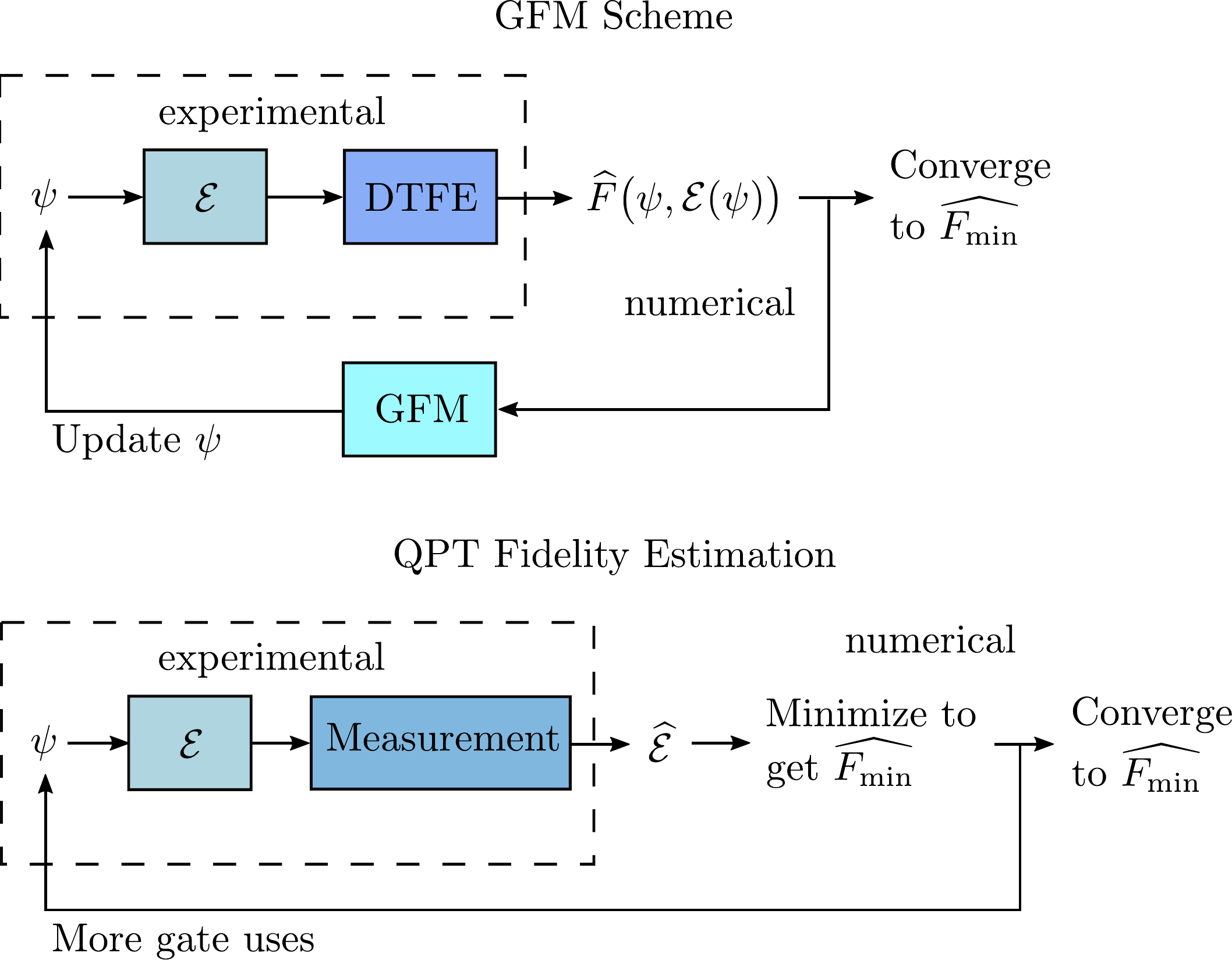}
\caption{\label{fig:scheme}Schematic depictions of the two approaches explored in our paper: GFM scheme and QPT fidelity estimation. Both approaches involve interwoven experimental and numerical steps. In the GFM approach, the experimental---via the direct target fidelity estimation procedure---and the numerical---via gradient-free descent---steps alternate seamlessly in a hybrid minimization algorithm, until convergence to the minimum fidelity $\widehat\Fmin$. In QPT fidelity estimation, the experimental reconstruction of the process matrix occurs first, followed by a numerical minimization to obtain $\widehat\Fmin$, and this is repeated, if needed, with more uses of the gate until the estimated accuracy of $\Fmin$ meets our set target.}
\end{figure}

Our scheme then proceeds as follows: We run, on a classical computer, a GFM algorithm to minimize $f(\psi)$. At each iterative step, the GFM suggests a set of trial states $\psi$, for which it requires the values $f(\psi)$. These values are obtained from the experiment by preparing those $\psi$ states and then performing the DTFE algorithm in the laboratory for each $\psi$. We iterate the GFM algorithm until a stopping rule is satisfied to confirm the attainment of a local minimum. The whole procedure is repeated sufficiently many times to find the global minimum with a specified (high) probability. The stopping rule is carefully tuned, as we will describe below, to attain the desired $\epsilon$ accuracy. The scheme is summarized in Fig. \ref{fig:scheme}.

\makeatletter\global\advance\@colroom10pt\relax\set@vsize\makeatother

In the following two sections, we describe, in greater detail, first the DTFE scheme of Ref.~\cite{DF1} and then the GFM algorithm employed in our numerical examples.

\subsection{Direct estimation of the target fidelity}\label{sec:DTFE}
We describe here the basic ideas of the DTFE scheme of Ref.~\cite{DF1} needed to understand our work. The analysis and formulas here are taken from that reference, and we refer the reader to the original paper for further details. The target fidelity, i.e., the quantity of interest in Ref.~\cite{DF1}, is the fidelity between a particular state $\rho$, imagined to be the actual state produced by a source, and a pure state $|\psi\rangle$, taken to be the target state the source is designed to produce in the ideal situation. The target fidelity hence quantifies how close the source is to that ideal. That the target state is pure is important technically for the linearity, in $\rho$ and $\psi$, of the squared fidelity used as the logical basis of the scheme; a pure target state is anyway the often-encountered situation in many quantum information processing tasks. For our purpose here, it suffices as well, as our minimization is over pure input states only.

We write the square of the target fidelity, for a $d=2^n$-dimensional situation, as
\begin{equation}\label{eq:fid1}
F(\psi,\rho)=\tr(\psi\rho)=\sum_kx_k(\psi)x_k(\rho),
\end{equation}
where $x_k(\,\cdot\,)\equiv \frac{1}{\sqrt d}\tr\bigl\{W_k(\,\cdot\,)\bigr\}$, with $W_k$ an $n$-qubit product Pauli operator, so that the set $\bigl\{\frac{1}{\sqrt d}W_k\bigr\}_{k=1}^{d^2}$ is an orthonormal operator basis. $x_k(\tau)$ for any state $\tau$ is then the (real) coefficient of $\frac{1}{\sqrt d}W_k$ when writing $\tau$ as a linear combination of elements of that Pauli operator basis. The sum over $k$ in Eq.~\eqref{eq:fid1} above is understood to be only over those $k$ values with $x_k(\psi)\neq 0$. In our present situation, the $x_k(\psi)$s are known, while the $x_k(\rho)$s are not.

The fidelity can be rewritten as
\begin{equation} \label{eq:fid2}
F(\psi,\rho)=\sum_k p_kX_k,
\end{equation}
with $p_k\equiv [x_k(\psi)]^2$ and $X\equiv\frac{x_k(\rho)}{x_k(\psi)}$. Note that $\{p_k\}$ is a probability distribution: $\sum_kp_k=F(\psi,\psi)=1$, and $p_k>0$. Let us define a random variable $X$ which takes value $X_k$ with probability $p_k$. Then, observe that $F(\psi,\rho)=\langle X\rangle$, the expectation value of $X$, which depends on the unknown $\rho$. Now, for each $k$, $x_k(\rho)$ can be estimated from the experiment by measuring $W_k$, a product-Pauli measurement, on $\rho$. Then, the target fidelity, now understood to be equal to $\langle X\rangle$, can be estimated by repeated trials where $W_k$ is chosen as the measurement to be performed on $\rho$ with probability $p_k$.

How good is this estimate? Suppose we do $h\equiv 1/(\eta\delta^2)$ trials, with $k_1,k_2,\ldots, k_h$ chosen according to the distribution $\{p_k\}$, and so obtain estimates for $X_{k_1},X_{k_2},\ldots,X_{k_h}$. Then, $Y\equiv \frac{1}{h}\sum_{\ell=1}^hX_{k_\ell}$ satisfies the inequality: $\textrm{Probability}{\left(|Y-\langle X\rangle|\geq\eta\right)}\leq \delta$.
We cannot, however, know $X_{k_\ell}$ precisely with only a finite number of copies of $\rho$. To estimate $X_{k_\ell}$, it suffices to measure $W_{k_\ell}$ on $t_{k_\ell}$ copies of $\rho$, where
\begin{equation}
t_{k_\ell}\equiv \frac{2\log (2/\delta)}{dp_{k_\ell}h\eta^2}.
\end{equation}
Then, $Y$, now built from estimates of $X_{k_\ell}$ using the above number of copies, satisfies instead the inequality
\begin{equation}
\textrm{Probability}{\left(|Y-\langle X\rangle|\geq2\eta\right)}\leq 2\delta.
\end{equation}
The expected total number of copies of $\rho$ required, to attain an estimate of the target fidelity $2\eta$ away from the true value is then
\begin{equation}
h\sum_kp_kt_k= \sum_k\frac{2\log (2/\delta)}{d \eta^2}\leq 2\log(2/\delta)\frac{d}{\eta^2},
\end{equation}
giving a scaling of $O(d/\eta^2)$ copies, for fixed $\delta$. A more accurate estimate of the expected number of copies is given in Ref.~\cite{DF1} by taking into account that $t_{k_\ell}$ has to be an integer, but that does not change the $O(d/\eta^2)$ scaling with $d$.

\subsection{Gradient-free minimization with CMA-ES}\label{CMAES}
Our GFM direct estimation scheme accommodates the use of any GFM algorithm. The efficiency of the GFM algorithm is of crucial importance in the present context, but other considerations such as ease of coding and number of tuning parameters can also affect one's choice.
There are many known GFM algorithms, including the downhill simplex method \cite{Simplex,ComplexitySimplex}, the directional direct search \cite{dds1,dds2,dds3}, the stochastic method \cite{GF5,SG1}, and the covariance matrix adaptation evolution strategy \cite{CMAES1,CMAES2,CMAES3,CMAES4} (CMA-ES), among others. The downhill simplex method is arguably the most well-studied one, but may not perform well in high-dimensional problems \cite{Torczon1989}. Trying out a few methods on our problem, we found CMA-ES to work well, converging more quickly than the downhill simplex method in our problem instances, with more stable performance than the directional direct search, and having fewer tuning parameters than the stochastic method. We thus focus on CMA-ES in our numerical examples below. We emphasize that a different user can choose a different GFM algorithm and observe a different efficiency performance; therein lies the potential for improvement beyond what we report here, with more efficient GFM methods.

Here, we provide some pertinent details of the CMA-ES algorithm, referring the reader to the original papers (Refs.~\cite{CMAES1,CMAES2,CMAES3,CMAES4}) for further explanation. We pay special attention to how we choose the stopping criteria that terminates the CMA-ES algorithm, as they are crucial for attaining the desired accuracy for $\Fmin$.

To implement the CMA-ES algorithm, we parameterize the domain space, the $d$-dimensional Hilbert space of pure ($n$-qubit) states, as $\psi=\ell \ell^\dagger/\tr(\ell\ell^\dagger)$, where $\ell$ is the complex column vector
\begin{equation}
\ell=(\ell_1+\upi \ell_2, \ell_3+\upi \ell_4,\ldots, \ell_{2d-1}+\upi \ell_{2d})^\mathrm{T}.
\end{equation}
The domain space is $2d$-dimensional, but there is a single extra parameter---the trace of $\psi$, constrained to be 1---when regarded as a parameterization for $\psi$. This extra parameter does no harm to the minimization, and, in fact, we observe this $2d$-parameter approach to work better for CMA-ES in our numerical examples than an alternative $(2d-1)$-dimensional parameterization with spherical coordinates.

In the $k$th iterative step of the CMA-ES algorithm, a set of $\lambda$ points in the domain space, $L^{(k)}\equiv{\left\{\ell_a^{(k)}\equiv\bigl(\ell_{a,1}^{(k)},\ell_{a,2}^{(k)},\cdots,\ell_{a,2d}^{(k)}\bigr)^\mathrm{T}\right\}}_{a=1}^\lambda$, is drawn from the normal distribution,
\begin{equation}
\ell_a^{(k)} \sim m^{(k)} + \sigma^{(k)}\mathcal{N}{\bigl(0,C^{(k)}\bigr)}, \quad \text{for} \; a=1,...,\lambda,
\end{equation}
where $m^{(k)}$, $\sigma^{(k)}$, and $C^{(k)}$ are the mean, step-size, and covariance matrix, respectively, for the $k$th step. At the initialization step, $m^{(1)}$ is set as a Haar-random pure state, $C^{(1)}$ is set equal to the identity matrix, and $\sigma^{(1)}$ is set to be 0.3. $m^{(1)}$ is set to be a Haar-random pure state to reflect our lack of knowledge of the state which attains the minimum value of fidelity. We observed empirically that the performance remains roughly the same even if we increase or decrease $\sigma^{(1)}$ by an order of magnitude. As long as $\sigma^{(1)}$ is not set to be too large or too small, the performance will not be affected too much. For the case with many local minima, choosing a larger value of $\sigma^{(1)}$ might increase the chance of finding the global minimum. The fidelity value $f_a^{(k)}\equiv f\bigl(\psi_a^{(k)}\bigr)$, for $\psi_a^{(k)}$ built from $\ell_a^{(k)}$, is estimated using the DTFE scheme for each $\ell_a^{(k)}\in L^{(k)}$. The points are then ranked according to the $f_a^{(k)}$ values, with $\ell_{a:\lambda}^{(k)}$ having the $a$th smallest value of $f$, i.e., $f\bigl(\psi_{1:\lambda}^{(k)}\bigr)\leq f\bigl(\psi_{2:\lambda}^{(k)}\bigr)\leq\cdots\leq f\bigl(\psi_{\lambda:\lambda}^{(k)}\bigr)$. The mean for the normal distribution is then updated by the weighted average,
\begin{equation}\label{eq:update}
m^{(k+1)} = \sum\limits^{\mu}_{a=1}w_a \ell^{(k)}_{a:\lambda},
\end{equation}
where $w_k>0$ and $\mu\leq\lambda$. The step-size and covariance matrix are also updated according to rules based on the ranking of the $f$ values. Further details of the algorithm, as well as the appropriate choice of parameters, can be found in Ref.~\cite{Hansen2016}.

In effect, the updates of $m^{(k)}$, $\sigma^{(k)}$, and $C^{(k)}$ move the ``region of interest'' within the domain space in the direction of smaller $f$ values, in correspondence with our goal of minimizing $f$. In the next iterative step, we draw the domain points $L^{(k+1)}$ from the updated region of interest and continue the move towards small $f$ values. This continues until the stopping criteria (more on these below) are met. Our estimate for $\Fmin$, for one run of CMA-ES, is then the fidelity for the best (smallest $f$) point found in the final iterative step, $\ell^{(\ldots)}_{1:\lambda}$. Each run of the CMA-ES algorithm returns an estimate of a local minimum value for $f$; the entire procedure is repeated sufficiently many times, with different initial conditions, to have confidence that one of those local minima is the global minimum. We estimate the number of repeats needed, for a $95\%$ chance of obtaining the global minimum, using random channels in our numerical simulation (see the numerical examples in Sec.~\ref{Examples}).

\begin{figure}[t]
\centering
\includegraphics[width=0.65\columnwidth]{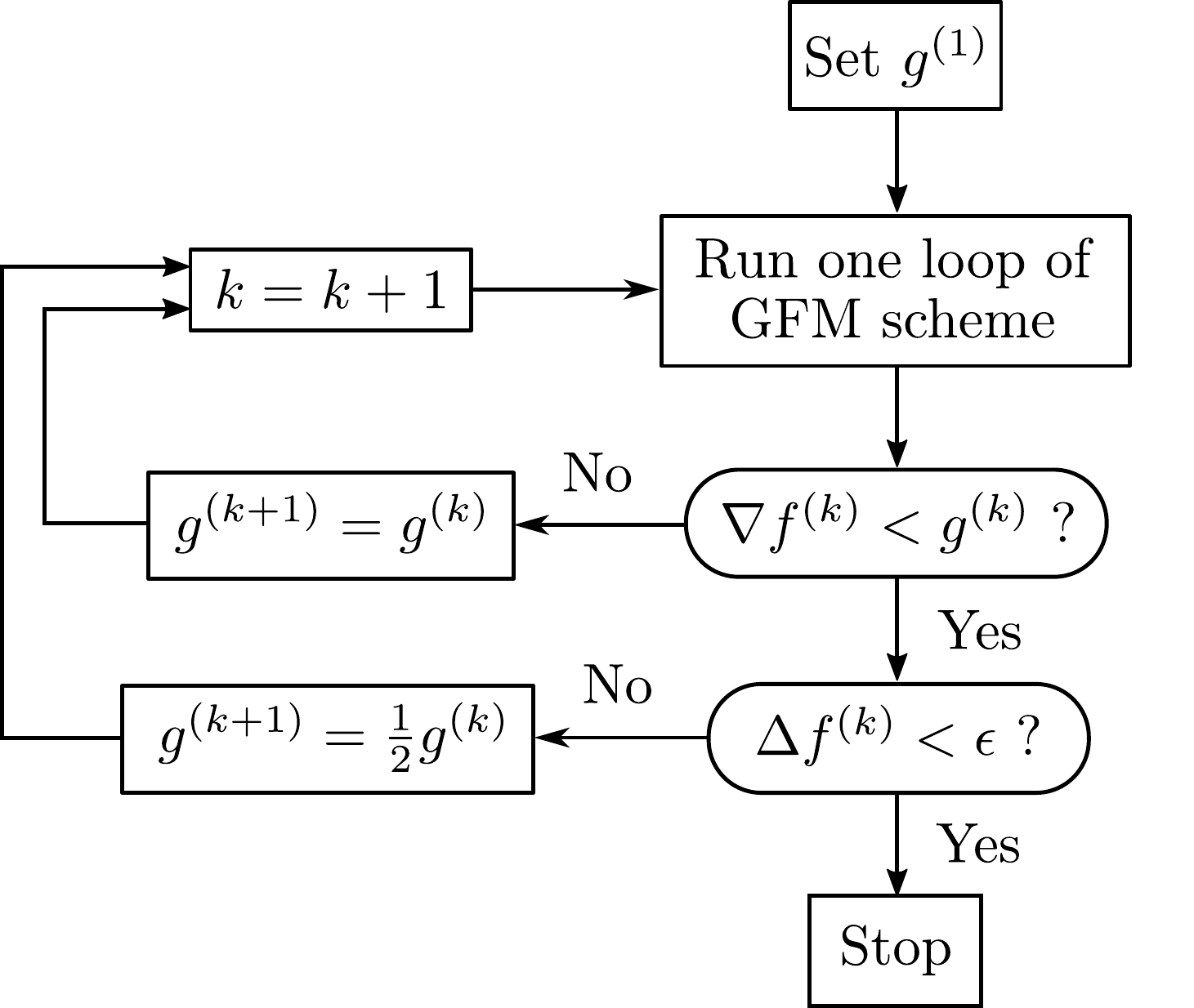}
\caption{ \label{fig:GFMStop} Diagrammatic representation of the stopping criteria used in the GFM scheme. See main text for a description of the various quantities.}
\end{figure}

To determine when to stop the CMA-ES descent, we calculate two quantities in each iterative step. The first is the range of fidelities for the current set of sample points $L^{(k)}$,
\begin{equation}
\Delta f^{(k)} \equiv \max\limits_a \{f^{(k)}_a\}-\min\limits_a \{f^{(k)}_a\}.
\end{equation}
The second is
\begin{equation}
\nabla f^{(k)} \equiv \frac{1}{\lambda}\sum\limits_{a=1}^\lambda \frac{|f^{(k)}_c - f^{(k)}_a|}{|\ell^{(k)}_c - \ell^{(k)}_a|},
\end{equation}
where $\ell^{(k)}_c$ is the centroid of $L^{(k)}$, and $f^{(k)}_c$ is the corresponding fidelity. $\nabla f^{(k)}$ can be taken as an estimation of the magnitude of the gradient at $\ell^{(k)}_c$.
The stopping criteria are then imposed adaptively to accommodate the variety of possible behaviors of the descent function $f(\psi)$ for different noisy channels. We set an initial threshold $g^{(1)}$ for the gradient estimate $\nabla f^{(1)}$. The target $\Fmin$ accuracy is $\epsilon$, as before. In subsequent iterative steps, if $\nabla f^{(k)}> g^{(k)}$, we retain the gradient threshold level, i.e., set $g^{(k+1)}=g^{(k)}$, and continue with the next iterative step. If, instead, $\nabla f^{(k)}<g^{(k)}$ but $\Delta f^{(k)}>\epsilon$, we halve the gradient threshold, i.e., set $g^{(k+1)} = g^{(k)}/2$, before taking the next iterative step. If $\nabla f^{(k)}<g^{(k)}$ and $\Delta f^{(k)}<\epsilon$, the algorithm is terminated. A schematic of the stopping criteria is given in Fig.~\ref{fig:GFMStop}.

The rationale for the adaptive stopping criteria is as follows. The algorithm should terminate when it gets to a stationary point, a minimum, where both the gradient and the range of fidelity values in the set of sample  points are small. That the range is small is demanded by our set target accuracy; that the gradient is small enough is needed to ensure that we have arrived at the minimum value. If we know the $f$ function, requiring the gradient alone to be small enough would also guarantee that the range is small, as the sample points are picked from a local region around the current position. However, since we do not have information about $f(\psi)$, we also do not know how small a gradient threshold is needed for sufficient convergence in the range. Instead, the gradient threshold has to be adjusted adaptively according to the noise channel at hand. We make use of the range of the fidelity values to judge whether the current gradient threshold is sufficiently stringent to ensure that we have arrived at the minimum point. If the gradient is below the set threshold, but the range is still beyond $\epsilon$, this indicates that our gradient criterion is simply not stringent enough; $g^{(k)}$ is thus halved and the iteration continues.

Note that the value of initial gradient threshold $g^{(1)}$ has to be pre-chosen according to the expected channel distribution and the target $\epsilon$ value, to prevent early termination of the algorithm. In our numerical examples, the value of $g^{(1)}$ was chosen, for each distribution of $n$-qubit channels, so that the probability (over the channel distribution) that $|\widehat{F_\text{min}}-F_\text{min}|\leq \epsilon$ is around $0.95$, with $\epsilon=0.01$, with the $0.95$ judged from numerical simulations using a number of trial true channels.

There is one more important detail that has to be addressed: the accuracy of the DTFE procedure needed for the CMA-ES algorithm, as quantified by the $\delta$ and $\eta$ parameters of Sec.~\ref{sec:DTFE}. The DTFE, with a given number of copies of the noisy state, estimates $f(\psi)$ only to a certain accuracy. If the DTFE accuracy is set too high, the resource cost of our GFM direct estimation scheme will become very high; if it is set too low, the CMA-ES algorithm may not converge to the correct minimum value, if it converges at all. CMA-ES, like many evolutionary algorithms of a similar flavor, is relatively robust to noise in function evaluations \cite{Arnold2002}, but proper handling of the accuracy of the function evaluation is still needed.  In our numerical examples, we set $\delta=0.05$ in our DTFE sub-routine and update the needed $\eta$ value in the $k$th iteration of the CMA-ES according to the uncertainty handling algorithm \cite{uncertaintyhandle,uncertaintyhandle2} described next.

The CMA-ES update rule of Eq.~\eqref{eq:update} is based on the ranking of the $f$ values. This means that noisy function evaluation will not affect it as long as the inaccuracies are not large enough to change the $f$ ranking. One can judge whether the function evaluation is sufficiently accurate by checking for rank changes after reevaluations of the function. In our specific context, reevaluation of $f(\psi)$ refers to repeating the DTFE procedure with the same parameters, to get a second estimate of the value of $f(\psi)$ for a given target state $\psi$. The inherent randomness of the DTFE procedure will yield different values, with larger variations from fewer number of uses of the channels, and hence more inaccurate evaluation of $f$.
We follow the uncertainty handling algorithm proposed in \cite{uncertaintyhandle,uncertaintyhandle2}. At each iterative step $k$, we reevaluate $f(\psi_a)$ for each $\psi_a$ built from $\ell_a^{(k)}\in L^{(k)}$, calculate the rank changes after the reevaluations, compute a measure of the uncertainty level, and adjust the accuracy of the DTFE procedure as needed. Specifically, we carry out the following steps:

\begin{center}
\textbf{Uncertainty Handling Algorithm}
\end{center}
\begin{enumerate}
\item For each $L^{(k)}$, compute $f$ for each $\ell_a^{(k)}\in L^{(k)}$ twice. Denote the two values obtained as $f_a^{(k)}$ and $\widetilde{f}_a^{(k)}$.
\item For each $a=1,...,\lambda$, compute the rank change $\Delta R_a^{(k)} = |R(f_a^{(k)})-R(\widetilde{f}_a^{(k)})|-1$, where $R(f_a^{(k)})$ and $R(\widetilde{f}_a^{(k)})$ are the ranks of $f_a^{(k)}$ and $\widetilde{f}_a^{(k)}$, respectively, in the combined set $\{f_a^{(k)},\widetilde{f}_a^{(k)}\}_{a=1}^\lambda$.
\item Compute the uncertainty level,
\begin{align}
\qquad s^{(k)}\!&\equiv\frac{1}{\lambda}\sum_{a=1}^\lambda \Bigl[2\Delta R_a^{(k)}\!-\!\pi_\theta\Bigl(\!R\bigl(\widetilde{f}_a^{(k)}\bigr)\!-\mathrm{H}\bigl(\widetilde{f}_a^{(k)}\!\!-\!f_a^{(k)}\bigr)\!\Bigr)\nonumber\\
&\quad\qquad -\pi_\theta\Bigl(R\bigl(f_a^{(k)}\bigr)-\mathrm{H}\bigl(f_a^{(k)}-\widetilde{f}_a^{(k)}\bigr)\Bigr)\Bigr],
\end{align}
where $\mathrm{H}(\cdot)$ is the step function and $\pi_\theta(R)$ is the $(50\theta)$-th percentile of all possible rank changes (given by the set $\{|1-R|,|2-R|,...,|2\lambda-1-R|\}$) for a given rank $R$.
\item If $s^{(k)}>0$, increase the accuracy in DTFE in the $(k+1)$-th iteration by setting $\eta^{(k+1)} = \alpha\eta^{(k)}$, with $0<\alpha<1$. If $s^{(k)}<0$, set $\eta^{(k+1)} = \eta^{(k)}/\alpha$.
\item For $a=1,...,\lambda$, set $f_a^{(k)} = \tfrac{1}{2}\bigl(f_a^{(k)}+\widetilde{f}_a^{(k)}\bigr)$.
\end{enumerate}
In our numerical examples, $\theta=0.7$ and $\alpha=\frac{1}{\sqrt{2}}$ were found to be good choices.

\section{QPT Fidelity Estimation}\label{sec:QPT}
The performance of our GFM direct estimation scheme has to be compared with the standard alternative of quantum process tomography (QPT), whereby the full process matrix of the unknown channel $\cE$ is first estimated, and then the $\Fmin$ value is numerically computed from the obtained process matrix. For completeness, we remind the reader of a few aspects of QPT important for our work; of course, many textbooks and papers are available on the subject (see, for example, Ref.~\cite{LNP649}).

QPT attempts to reconstruct the full description of an unknown quantum channel (or process) $\cE$ from a finite number of uses of the channel. A chosen set of states $\{\rho_k\}$ is sent in as inputs to the channel, and state tomography is done on the outputs of the channel, using a measurement (a positive operator-valued measure, or POVM) $\{\Pi_\ell\}$. The probability of getting outcome $\Pi_\ell$ if the input state $\rho_k$ was sent is given by the Born rule,
\begin{equation}\label{eq:BornRule}
p_{k\ell}=\tr\bigl(\Pi_\ell\cE(\rho_k)\bigr).
\end{equation}
One estimates the values of the $p_{k\ell}$s from the experiment using the channel $N_{k}$ times for each input state $\rho_k$, amounting to a total of $N=\sum_{k}N_k$ uses of the channel. The sets $\{\rho_k\}$ and $\{\Pi_\ell\}$ are chosen to be informationally complete, i.e., there is a one-to-one mapping between the $p_{k\ell}$s and the quantum channel $\cE$. Once we have the estimate for $\{p_{k\ell}\}$, we apply the mapping to get an estimate for $\cE$.

In our numerical examples below, we used a particular choice of $\{\rho_k\}$ and $\{\Pi_\ell\}$, namely, the product tetrahedron states and the product tetrahedron measurement. The set of $n$-qubit product tetrahedron states is the pure states $\psi_{k_1,k_2,\ldots,k_n}\equiv\psi_{k_1}\otimes\psi_{k_2}\ldots\otimes\psi_{k_n}$, with each $\psi_k$ a single-qubit state written in the Bloch-sphere representation as
\begin{equation}
\psi_k=\tfrac{1}{2}(\id+\bm{a}_k\cdot\bm{\sigma}),\qquad k=1,2,3,\textrm{and }4,
\end{equation}
where the four $\bm{a}_k$s are three-dimensional unit vectors subtending a tetrahedron in the Bloch sphere \cite{Tetrahedronpaper}, and $\bm{\sigma}$ is the vector of Pauli operators. The product tetrahedron measurement is also defined in terms of these tetrahedron states: $\Pi_{\ell_1,\ell_2,\ldots, \ell_n}\equiv \Pi_{\ell_1}\otimes\Pi_{\ell_2}\ldots\otimes\Pi_{\ell_n}$, with each $\Pi_\ell\equiv \frac{1}{2}\psi_\ell$, for $\ell=1,2,3,$ and 4. The tetrahedron states and measurements form an informationally complete QPT scheme. For $n$ qubits, the scheme requires $4^{2n}$ different settings of input state and output measurement. Owing to the tetrahedron geometry, this scheme uses the minimal number of different input states and measurement outcomes in a symmetric way.

The data from the QPT experiment consist of a sequence of detector clicks, which we summarize into $\{n_{k\ell}\}$, with $n_{k\ell}$ being the number of clicks in the detector for $\Pi_\ell$ for input state $\rho_k$, where $k,\ell=1,2,\ldots,4^n$. The total number of uses of the channel is $N= \sum_{k\ell}n_{k\ell}$. To reconstruct the channel $\cE$ from the data, we first do linear inversion, i.e., solving for $\cE$ (using the Choi-state representation) by replacing, in the Born rule given by Eq.~\eqref{eq:BornRule}, $p_{k\ell}$ with the relative frequency $n_{k\ell}/N$. This does not guarantee that the $\cE$ obtained from the linear inversion will be a valid, i.e., completely positive, channel. Complete positivity is then enforced by projecting the solution from linear inversion onto the nearest CPTP map using the algorithm of Ref.~\cite{CPTPalgorithm}, thereby giving us an estimate $\widehat \cE$ of the channel. The minimum fidelity $\Fmin$, our quantity of interest, is then estimated by numerically minimizing $F(\psi,\widehat{\cE}(\psi))$ over $\psi$, with standard conjugate-gradient methods.

We want to obtain an estimate of $\Fmin$ that is within $\epsilon$ of the true value. The accuracy of $\Fmin$ is controlled by the accuracy of the estimate $\widehat{\cE}$ --- a better estimate of $\cE$, obtained with a larger $N$, will give a more accurate estimate of $\Fmin$. \textit{A priori}, we do not know the $N$ needed to attain the specified accuracy on $\Fmin$. That depends on the unknown channel $\cE$. For a fair comparison with our GFM scheme, we need $N$ just large enough so that $\Fmin$ attains the desired accuracy. We thus increase $N$ slowly and look for convergent behavior. See Fig.~\ref{fig:scheme} for the summary of the scheme.

\begin{figure}[t]
\centering
\includegraphics[width=0.65\columnwidth]{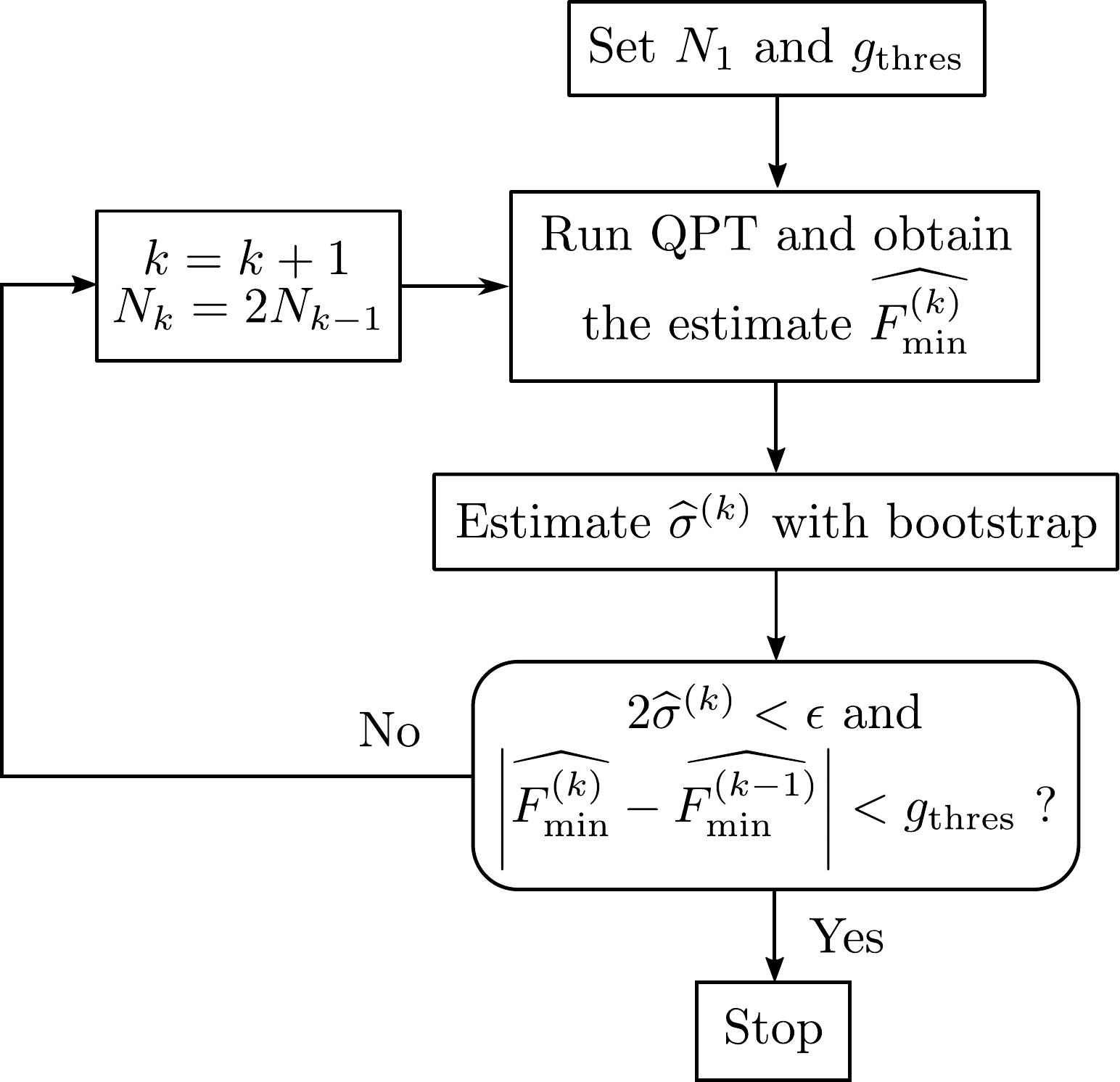}
\caption{ \label{fig:QPTStop} Diagrammatic representation of the stopping criteria used in the QPT fidelity estimation. See main text for a description of the various quantities.}
\end{figure}

We begin with a small number of uses of the channel $N_1$ to estimate $\cE$ and then $\Fmin$. In the next iteration, we double the number of uses by measuring a further $N_1$ uses, combining the obtained data with that from the previous iteration to obtain a second estimate of $\Fmin$. In subsequent iterations, the amount of data used to estimate $\Fmin$ is doubled each time by doubling the number of uses of the channel. We continue the iteration, obtaining (hopefully) more and more accurate estimates of $\Fmin$ until both stopping criteria are met. The first requires that the error bar $\widehat\sigma^{(k)}$ ---estimated through bootstrapping of the obtained data---of $\Fmin$ is smaller than $2\epsilon$. The second criterion demands that the change of $\Fmin$ from the previous iteration (with half the number of channel uses) is smaller than some threshold $g_\mathrm{thres}$. A schematic of the stopping criteria is given in Fig.~\ref{fig:QPTStop}.

Note that despite the similarity to our GFM approach where the procedure was also terminated according to range and gradient criteria, the reason why a gradient criterion is needed here for QPT is rather different. The $\Fmin$ estimation from QPT was observed in our numerical investigations to be biased, usually underestimating the $\Fmin$ unless $N$ is large enough. Similar to how we chose our $g^{(1)}$ values for the GFM scheme, suitable values of $g_\mathrm{thres}$, different for different number of qubits and $\epsilon$ values, were chosen by running tests for 100 randomly chosen trial channels and setting $g_\mathrm{thres}$ such that the $\Fmin$ value is within the desired accuracy for 95\% of the test channels.

\section{Performance of our scheme}\label{sec:perf}

How well does our GFM scheme perform, in terms of the number of uses of the channel, compared with the alternative using QPT fidelity estimation? We are interested, in particular, with the scaling of the resource cost with the number of qubits, for fixed target precision $\epsilon$. The resource cost of QPT is well known to scale poorly with the dimension; does our scheme do better in that respect?

Intuitively, one would expect our GFM scheme to outperform QPT fidelity estimation. Our scheme directly estimates the one quantity of interest, $\Fmin$, while QPT first estimates the full process matrix, which contains all information about the channel, and then estimates $\Fmin$, discarding the rest of the gathered information. Our minimalistic approach should thus win over QPT. Indeed, resource estimates suggest this, as we explain below. However, our numerical examples, which take into full account the complexities of both schemes, as well as the particular class of channels considered, tell a somewhat different story. Below, we first discuss resource estimates to give some indication of potential performance and then describe the numerical comparison obtained from simulations of two specific classes of noise channels.

\subsection{Resource estimates}\label{sec:theory}

We can gain some insights into the possible performance by considering known theoretical bounds on the various components of our scheme as well as on QPT. We begin with QPT. A general scaling law for the resource cost of quantum \emph{state} tomography was derived in Refs.~\cite{Kueng2014,SampleComplexity}: $O(dr^2/\epsilon'^2)$ number of copies are needed to estimate the density operator $\rho$ of a $d$-dimensional system to accuracy $\epsilon'$, as measured by the trace-distance deviation from the true state. $r$ is the rank of $\rho$. The channel-state duality via the Choi--Jamio{\l}kowski isomorphism \cite{choipaper,Jamio} allows us to apply this result directly to QPT. Specifically, a $d$-dimensional quantum channel corresponds to a $d^2$-dimensional Choi state. Further assuming that the Choi state is full rank (supposing we have no reason to assume otherwise), we see that the bound for the resource cost of QPT becomes $O(d^6/\epsilon'^2)$ uses of the quantum channel, to estimate the Choi state within $\epsilon'$ \mbox{(trace)} distance from that of the true channel.

We are, however, interested in the accuracy of $\Fmin$, not in the Choi-state trace distance. Accurate reconstruction of the Choi state of the channel, of course, assures that the estimate of $\Fmin$ will be accurate as well. 
Specifically, we can show a relationship between the deviation in minimum fidelity and the Choi-state trace distance (see Appendix~\ref{sec:app1}),
\begin{equation}\label{eq:fidTr}
|\Fmin'-\Fmin|\leq 2d\Vert\rho_{\cE'}-\rho_\cE\Vert_\mathrm{tr},
\end{equation}
for the Choi states $\rho_{\cE}$ and $\rho_{\cE'}$ of two arbitrary channels $\cE$ and $\cE'$, respectively, where $\Fmin$ and $\Fmin'$ are the minimum fidelity of the channels $\cE$ and $\cE'$, respectively, and $\Vert M\Vert_\tr\equiv \frac{1}{2}\tr\bigl(\sqrt{M^\dagger M}\bigr)$ denotes the trace norm. The factor of $d$ in the first term on the right-hand side of Eq.~\eqref{eq:fidTr} is unavoidable; in fact, there exist (see Appendix~\ref{sec:app1}) one-parameter families of channels $\cE$ and $\cE'$ such that $|\Fmin'-\Fmin|= d\Vert\rho_{\cE'}-\rho_\cE\Vert_\mathrm{tr}$.

To attain accuracy $\epsilon$ for $\Fmin$, i.e., $|\widehat{\Fmin}-\Fmin|\leq \epsilon$, Eq.~\eqref{eq:fidTr} indicates that it is sufficient that the trace-distance accuracy of the QPT scheme satisfies $\Vert\widehat{\rho_\cE}-\rho_\cE\Vert_\mathrm{tr} \leq \epsilon'$, where $\epsilon'=\epsilon/(2d)$ and $\widehat{\rho_\cE}$ is the reconstructed Choi state from QPT. This yields a resource cost of $O(d^6/(\epsilon/d)^2)=O(d^8/\epsilon^2)$, i.e., $O(d^8)$ uses of the channel for fixed $\epsilon$, an altogether prohibitive scaling.

Next, we look at the resource estimates for our GFM scheme. Our two main ingredients, the DTFE scheme and the chosen GFM algorithm, directly determine the resource cost. The efficiency of the chosen GFM algorithm determines the number of function evaluations needed to arrive at the minimum point; the efficiency of the DTFE scheme determines how many uses of the channel are needed for each function evaluation to obtain an estimate of the required accuracy.
For the latter, Ref.~\cite{DF1} gives $O(d/\eta^2)$ for estimating the target fidelity to an accuracy $\eta$. We cannot, however, directly use this estimate. As explained earlier, in our GFM scheme, $\eta$ is not a fixed value, but is a quantity that is adjusted as the GFM iteration proceeds. We thus also need to consider possible dimensional dependence of $\eta$. In all our numerical examples, we observe that $\eta\sim 1/d$, for the $\eta$ values used in the course of the iterations, for $\Fmin$ accuracy $\epsilon=0.01$. Using this empirical estimate, we thus have that each use of the DTFE subroutine requires $O(d^3)$ uses of the channel.

For a GFM algorithm, there are generally two aspects to consider when estimating the resource cost: (i) the number of function evaluations needed in the course of the descent to a local minimum; (ii) the number of repeats of the descent to obtain the global minimum with high confidence. In our numerical examples fixed at $\epsilon=0.01$, we observe that only a constant, $d$-independent number of repeats was needed to arrive at the global minimum with high confidence~\cite{FN}. We thus disregard aspect (ii) in our considerations of the resource cost. For aspect (i), the CMA-ES algorithm has been observed empirically \cite{CMAES3} to require $O(d)$ to $O(d^2)$ function evaluations in total, to attain a fixed accuracy of the extremum function value. Combining this with the fact that each function evaluation invokes the DTFE scheme that requires $O(d^3)$ uses of the channel, we see that our scheme requires $O(d^4)$ to $O(d^5)$ channel uses, a possible improvement over QPT fidelity estimation.

\subsection{Numerical examples}\label{Examples}
While resource estimates can provide initial clues to the performance of a scheme, a more accurate test comes from numerical simulations of the scheme. Numerical tests, in particular, are able to give indications of variations in performance for different classes of channels, an aspect often not easily captured in a theoretical analysis.
In this section, we present our numerical comparisons of the performance of our GFM scheme with that of QPT fidelity estimation to estimate the minimum fidelity for channels on one to five qubits.

\begin{figure}
\centering
\includegraphics[width=\columnwidth]{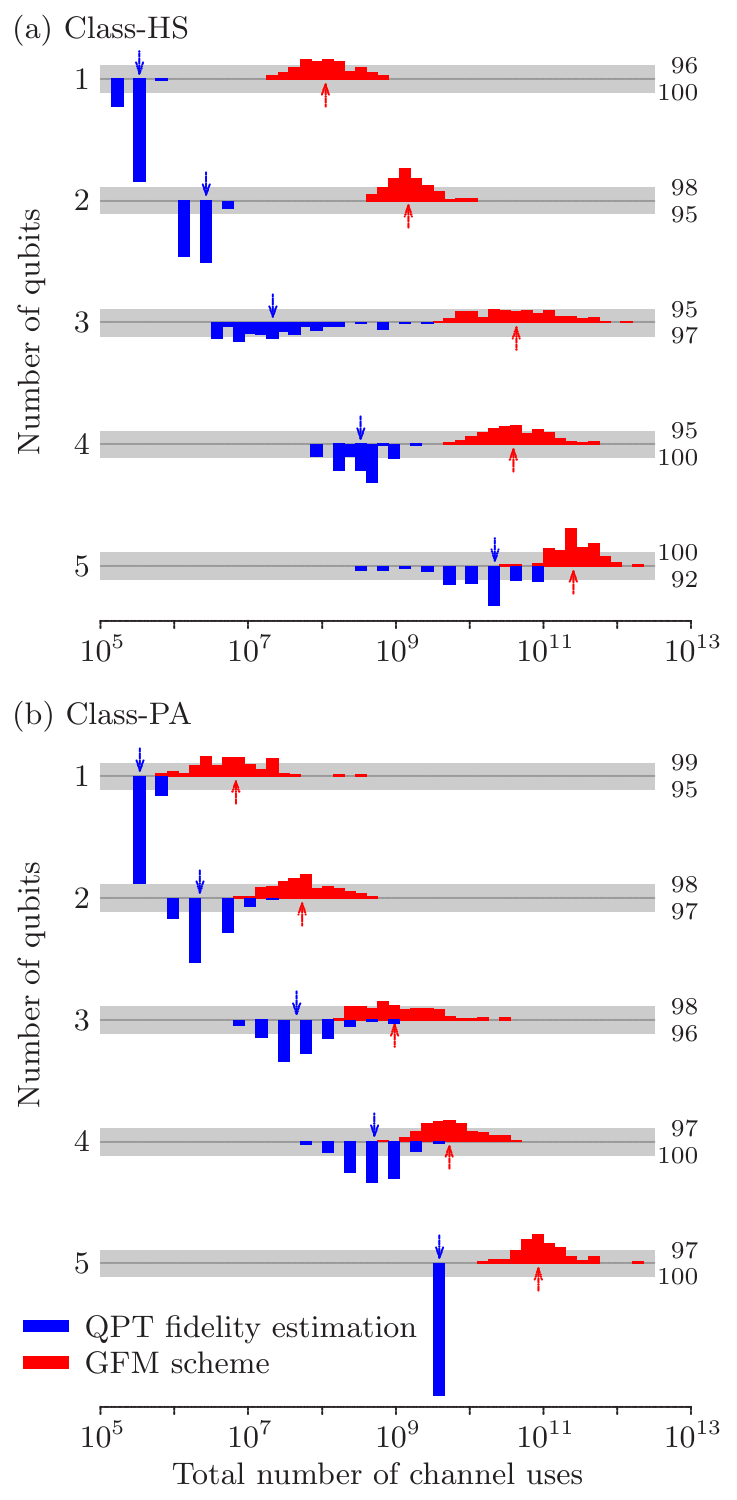}
\caption{\label{fig:numerics}The total number of channel uses for our GFM scheme and for QPT fidelity estimation for $n$ qubits and fixed target $\Fmin$ accuracy of $\epsilon=0.01$. Two classes of random channels were used: (a) Class-HS, and (b) Class-PA (see main text). Each histogram is generated from 100 randomly chosen channels, with the same 100 channels used for both schemes. The medians are indicated by the arrows; the gray bars mark the height of 10 counts. The numbers on the right (top number for GFM, bottom for QPT) of the histograms give the number of channels, out of 100, for which the estimated $\Fmin$ is within the desired accuracy. The stopping criteria are adjusted so that these numbers are around 95. The scaling of the number of channel uses with the dimension $d=2^n$ is observed, using the medians of the histograms, as (a) Class-HS: $O(d^{4.2})$ for QPT and $O(d^{2.7})$ for GFM; and (b) Class-PA: $O(d^{3.8})$ for QPT and $O(d^{3.4})$ for GFM.}
\end{figure}

The test channels used in our numerical simulations are drawn from two commonly encountered classes of random channels. It is possible to construct specific channels where one of the schemes outperforms the other. However, these examples are artificial and seldom relevant in practice. The first class, which we label as Class-HS, comprises channels chosen randomly according to the Hilbert--Schmidt measure on the space of quantum channels \cite{randomChannels}. To sample from this class, Haar-random unitary matrices $U$ are chosen on a $d^2$-dimensional Hilbert space, considered as the space of the system with an equal-dimensional ancilla A, and the channel is defined as $\cE(\cdot)\equiv \tr_\mathrm{A}\bigl(U(\cdot\otimes\psi_\mathrm{A})U^\dagger\bigr)$, for $\psi_\mathrm{A}$ a pure state on A. This first class of channels is a well-studied class used in many discussions of generic properties of quantum channels.

The second class, which we label as Class-PA, imitates noise channels observed in many quantum experiments today: $n$-qubit weak noise channels composed of a random Pauli channel followed by a random amplitude-damping channel. Here, the Pauli channel is $\cE_\mathrm{Pauli}(\cdot)\equiv\sum_{k=1}^{d^2}p_k W_k\rho W_k$, for $d\equiv 2^n$, with $W_k$ a tensor-product Pauli operator and $\{p_k\}$ a probability distribution. The random Pauli channel is generated by drawing numbers $q_2,q_3,\ldots, q_{d^2}$ uniformly from a $(d^2-2)$-simplex, and then setting $p_k=uq_k$, for $k=2,3,\ldots,d^2$, with $u$ uniformly randomly chosen from the range $0$ to $0.1$, and $p_1\equiv 1-\sum_{k=2}^{d^2}p_k$. The upper limit of $0.1$ on $u$ ensures that we have a weak (i.e., close to the identity) Pauli channel. The amplitude-damping channel is $\cE_\mathrm{AD}(\cdot)\equiv \sum_{j=0}^{d-1} E_j (\cdot)E_j^\dagger$, with
\begin{align}
E_0&\equiv \sum_{j=0}^{d-1}\sqrt{1-\gamma_j}|\psi_j\rangle\langle\psi_j|,\nonumber\\
\textrm{and} \quad E_j&\equiv \sqrt{\gamma_j} |\psi_0\rangle\langle \psi_j|, \quad j=1,2,\ldots,d-1,
\end{align}
where $\gamma_0= 0$, $\gamma_1= 0.1$, $\gamma_2,\gamma_3,\ldots \gamma_{d-1}$ are uniformly randomly chosen from the range $0$ and $0.1$, and $\{\psi_j\}_{j=0}^{d-1}$ is a Haar-random basis for the system Hilbert space. The amplitude-damping channel models population decay of the $n$-qubit system towards some $|\psi_0\rangle$ state. Physically, $|\psi_0\rangle$ is typically the energetic ground state of the system, which may or may not be aligned with the chosen Pauli axes directions.

In each numerical experiment, we randomly (from the chosen class of channels) choose an $n$-qubit channel and estimate $\Fmin$ using our GFM scheme and, separately, using QPT fidelity estimation. The target accuracy for $\Fmin$ is set to $\epsilon=0.01$. The experiment is repeated 100 times (i.e., 100 different channels) for a fixed channel class and fixed $n$. The tuning parameters, namely, the parameters in the CMA-ES algorithm and the DTFE procedure, and those in the stopping criteria for the GFM scheme and the QPT fidelity estimation scheme are prechosen for each channel class and $n$ (see Sec.~\ref{sec:directEst}) and fixed throughout the 100 experiments. The histograms for the experiments are given in Fig.~\ref{fig:numerics}.

The numerical experiments tell a rather different story than the resource estimates of the previous section. The most striking feature of Fig.~\ref{fig:numerics} is that QPT fidelity estimation requires fewer uses of the channel than our GFM scheme, at least for up to the tested five-qubit situation. In terms of scaling with the dimension of the system, the number of uses of the channel for our GFM scheme is about $O(d^3)$, not far from the resource estimates; that for QPT is, however, closer to $O(d^4)$, rather than the $O(d^8)$ behavior of the resource estimates. Overall, the numerics suggest that our GFM scheme shows some advantage over QPT in terms of scaling with the size of the system, with a larger advantage for Class-HS than for Class-PA (see Fig.~\ref{fig:numerics}). However, when the system size is small, i.e., the case of practical interest in the near future, QPT outperforms our scheme in actual number of uses of the channel.

One possible reason behind the significant difference between the observed numerical scaling and the resource estimates on the number of channel uses is that the inequality relation in Eq.~\eqref{eq:fidTr} merely provides an upper bound on the $\Fmin$ deviation. Given a true channel $\cE$ and its estimate $\widehat\cE$ from QPT, the deviation in minimum fidelity of $\cE$ and $\widehat\cE$ is typically much smaller than the upper bound indicated in Eq.~\eqref{eq:fidTr}. This is particularly true for the two classes of channels in our numerical examples, as we observed empirically. Therefore, the use of Eq.~\eqref{eq:fidTr} grossly overestimates how stringent we need to be in the trace-distance deviation to achieve a desired fidelity deviation. This highlights the importance of directly incorporating the desired figure of merit---the minimum fidelity in our case---into the QPT scheme, using it within the stopping criteria, rather than imposing a target accuracy on the deviation of the full process matrix.

Note that the spread in the number of channel uses visible in Fig.~\ref{fig:numerics} is not solely due to our random choice of channels. The fluctuations in the data (and hence the final estimated $\Fmin$ values) for each run of both schemes also contribute to the observed spread. The histograms shown in Fig.~\ref{fig:numerics} show only one run per scheme per random channel. Additional numerical studies indicate that if each procedure were repeated 100 times on the \emph{same} channel, the spread in the number of channel uses would be roughly as large as what is seen in Fig.~\ref{fig:numerics}. Also, note that the gaps in the histograms for the QPT fidelity estimation in Fig.~\ref{fig:numerics} (blue bars) are due to our choices of $N_1$, the base number of channel uses, subsequently doubled in each round of the iterative procedure.

A further remark concerns the resource scaling of our GFM scheme for Class-HS. Observe in Fig.~\ref{fig:numerics}(a) that we need $O(d^{2.7})$ channel uses for our scheme, a more favorable scaling than the $O(d^{4.2})$ of QPT fidelity estimation. In fact, we suspect a further slowdown in the increase in channel uses with $d$ beyond five qubits for our scheme, gaining further advantage over QPT fidelity estimation. This is because, numerically, we observe that for a typical channel from Class-HS, the range of fidelity values over all pure states shrinks logarithmically as $d$ increases. This means that for large $d$ and fixed target accuracy for the estimation of $\Fmin$, we only need to pick \emph{any} pure state $\psi$, estimate $f(\psi)$ using the DTFE scheme, and that is already close to the true $\Fmin$ value, even without further minimization using the GFM algorithm. We thus expect the number of uses of the channel for our GFM scheme to be much reduced in that case. Note that this reduction in the fidelity range is not observed for Class-PA.

As the classical computational resource remains one of the limiting factors of tomography, it is also important to compare the two schemes in this aspect. The classical computational resource cost of the GFM scheme is estimated to be $O(d^5)$. For the QPT fidelity estimation scheme, the estimated classical computational resource cost is at least $O(d^6)$. See Appendix~\ref{sec:app2} for the discussion. In this aspect, the GFM scheme has better performance compared with the QPT fidelity estimation scheme.

\section{Conclusion}\label{sec:conc}

We have explored two different direct schemes for estimating the minimum gate fidelity. The GFM scheme presents an interesting hybrid application of numerical GFM algorithms together with the experimental procedure of direct target fidelity estimation. We compared this to the alternative approach of QPT fidelity estimation, a key difference of which from standard QPT procedures is the direct incorporation of the quantity of interest as the stopping criterion in the iteration. Resource estimates suggest an extremely high cost for QPT fidelity estimation, with a scaling for $O(d^8)$ gate uses, compared with $O(d^4)$ for the GFM approach. This large difference in performance, however, was not seen for the numerical tests carried out on specific classes of noise channels, with both schemes showing a gate-use scaling closer to $O(d^4)$. This reminds us of the need to examine the performance of every procedure for the particular context at hand, in addition to general results that apply in all situations.

In practice, one could consider using either scheme. QPT fidelity estimation has the advantage that the QPT experiment can be easier to perform, using familiar measurement setups. Its performance, however, appears limited to $O(d^4)$ gate uses (for our specific numerical examples) to achieve an estimate of the minimum fidelity of a specified accuracy. On the other hand, while the current performance in our numerical examples is also $O(d^4)$, the GFM approach presents a potential for future improvement with an increase in the efficiency of the numerical GFM scheme used. The GFM approach is also more efficient in terms of classical computational resource cost. The downside, though, is that the GFM scheme, with the need for preparing arbitrary input states suggested by the GFM iteration,  can be more difficult to implement experimentally. That may, nevertheless, be well worth the effort if the number of gate uses can be significantly lowered for large $d$.

\vspace*{\fill}

\medskip
\begin{acknowledgements}
This work is supported by the Ministry of Education, Singapore (through Grant No. MOE2016-T2-1-130). The Centre for Quantum Technologies is a Research Centre of Excellence funded by the Ministry of Education and the National Research Foundation of Singapore.
\end{acknowledgements}

\vspace*{\fill}

\appendix
\section{Derivation of Eq.~(\ref{eq:fidTr})}\label{sec:app1}
Here, we explain the steps leading to Eq.~(\ref{eq:fidTr}) which relates $|\Fmin'-\Fmin|$ to $\Vert \rho_{\cE'}-\rho_\cE\Vert_{\tr}$.
We are concerned with two arbitrary CPTP channels, $\cE$ and $\cE'$. The Choi state of $\cE$ is denoted as $\rho_\cE\equiv(\id\otimes\cE)(\Phi)$, where $\Phi\equiv|\Phi\rangle\langle\Phi|$ with $|\Phi\rangle\equiv \frac{1}{\sqrt{d}}\sum_{i=1}^d|i\rangle|i\rangle$, a specific choice of a maximally entangled bipartite state. The fidelity of state $|\psi\rangle$ under the action of $\cE$ is $F_\cE(\psi)\equiv F(|\psi\rangle,\cE(\psi))=\langle\psi|\cE(\psi)|\psi\rangle$, and the minimum fidelity $\Fmin$ is attained by state $|\psi_\cE\rangle$. Analogous definitions apply for the channel $\cE'$, with Choi state $\rho_{\cE'}$ and minimum fidelity $\Fmin'$, attained by state $|\psi_{\cE'}\rangle$.\newpage

We begin with the fidelity difference, and employ the triangle inequality,
\begin{align}\label{eq:app1}
&\quad~|\Fmin'-\Fmin|={\left|F_{\cE'}(\psi_{\cE'})-F_\cE(\psi_\cE)\right|}\\
&=\tfrac{1}{2}{\left| F_{\cE'}(\psi_{\cE'})-F_\cE(\psi_{\cE'})+F_\cE(\psi_{\cE'})-F_\cE(\psi_\cE)\right.}\nonumber\\
&\qquad +{\left.F_{\cE'}(\psi_{\cE'})-F_{\cE'}(\psi_\cE)+F_{\cE'}(\psi_\cE)-F_\cE(\psi_\cE)\right|}\nonumber\\
&\leq\tfrac{1}{2}{\left| F_{\cE'}(\psi_{\cE'})-F_\cE(\psi_{\cE'})\right|}+\tfrac{1}{2}{\left|F_{\cE'}(\psi_\cE)-F_\cE(\psi_\cE)\right|}\nonumber\\
&\qquad +\tfrac{1}{2}{\left|F_{\cE'}(\psi_{\cE'})-F_{\cE'}(\psi_\cE)\right|}+\tfrac{1}{2}{\left|F_\cE(\psi_{\cE'})-F_\cE(\psi_\cE)\right|}\nonumber\\
&=\tfrac{1}{2}{\left| F_{\cE'}(\psi_{\cE'})-F_\cE(\psi_{\cE'})\right|}+\tfrac{1}{2}{\left|F_{\cE'}(\psi_\cE)-F_\cE(\psi_\cE)\right|}\nonumber\\
&\qquad +\tfrac{1}{2}{\left(F_{\cE'}(\psi_\cE)-F_{\cE'}(\psi_{\cE'})\right)}+\tfrac{1}{2}{\left(F_\cE(\psi_{\cE'})-F_\cE(\psi_\cE)\right)}\nonumber\\
&\leq \tfrac{1}{2}{\left| F_{\cE'}(\psi_{\cE'})-F_\cE(\psi_{\cE'})\right|}+\tfrac{1}{2}{\left|F_{\cE'}(\psi_\cE)-F_\cE(\psi_\cE)\right|}\nonumber\\
&\qquad +\tfrac{1}{2}{\left|F_{\cE'}(\psi_\cE)-F_\cE(\psi_\cE)\right|}+\tfrac{1}{2}{\left|F_{\cE'}(\psi_{\cE'})-F_\cE(\psi_{\cE'})\right|}\nonumber\\
&={\left| F_{\cE'}(\psi_{\cE'})-F_\cE(\psi_{\cE'})\right|}+{\left|F_{\cE'}(\psi_\cE)-F_\cE(\psi_\cE)\right|}.\nonumber
\end{align}
Above, we noted that $F_\cE(\psi_{\cE'})\geq F_\cE(\psi_\cE)$ and $F_{\cE'}(\psi_\cE)\geq F_{\cE'}(\psi_{\cE'})$ since $F_\cE$ and $F_{\cE'}$ attain their minimum values on $\psi_\cE$ and $\psi_{\cE'}$, respectively.

Now, straightforward calculation tells us that $F_\cF(\psi)=d\langle\overline\psi|\langle\psi|\rho_\cF|\overline\psi\rangle|\psi\rangle$ for any state $|\psi\rangle$, a CPTP channel $\cF$, and its associated Choi state $\rho_\cF$. Here, $|\overline\psi\rangle$ is a state associated with $|\psi\rangle$ by the Choi--Jamio{\l}kowski isomorphism (see, for example, Ref.~\cite{Sim2020}, Sec.~II for further explanation). Then, for any state $|\psi\rangle$, 
\begin{align}\label{eq:app2}
{\left|F_{\cE'}(\psi)-F_\cE(\psi)\right|}&=d\langle\overline\psi|\langle\psi|{\left(\rho_{\cE'}-\rho_\cE\right)}|\overline\psi\rangle|\psi\rangle\nonumber\\
&\leq d\max_{|\Psi\rangle}\langle\Psi|(\rho_{\cE'}-\rho_\cE)|\Psi\rangle,
\end{align}
We can relate this to the trace distance $\Vert\rho_{\cE'}-\rho_\cE\Vert_\mathrm{tr}=\frac{1}{2}\sum_i|\lambda_i|$, where $\lambda_i$s are the eigenvalues of $\rho_{\cE'}-\rho_\cE$. Since $\rho_{\cE'}-\rho_\cE$ is traceless, i.e., $\sum_i\lambda_i=0$, we know that $\sum_i\bigl|\lambda_i^{(+)}\bigr|=\sum_i\bigl|\lambda_i^{(-)}\bigr|$, where $\lambda_i^{(+)}$ and $\lambda_i^{(-)}$ are, respectively, the positive and negative eigenvalues of $\rho_{\cE'}-\rho_\cE$. Hence, $\Vert\rho_{\cE'}-\rho_\cE\Vert_\mathrm{tr}=\sum_i\bigl|\lambda_i^{(+)}\bigr|=\sum_i\bigl|\lambda_i^{(-)}\bigr|\geq \max_{|\Psi\rangle}\langle\Psi|(\rho_{\cE'}-\rho_\cE)|\Psi\rangle$, so that we have
\begin{equation}
{\left|F_{\cE'}(\psi)-F_\cE(\psi)\right|}\leq d\Vert\rho_{\cE'}-\rho_\cE\Vert_\mathrm{tr}.
\end{equation}
Applying this inequality to the two terms in the last line of Eq.~\eqref{eq:app1}, we find
\begin{equation}\label{eq:app3}
|\Fmin'-\Fmin|\leq 2d\Vert\rho_{\cE'}-\rho_\cE\Vert_\mathrm{tr},
\end{equation}
yielding the inequality of Eq.~\eqref{eq:fidTr}.

The factor of $d$ on the right-hand side is unavoidable: Let $\cD(\cdot)\equiv \tr(\cdot)\id/d$ denote the $d$-dimensional erasure channel, and let $\cD'(\cdot)\equiv\cD(\cdot)-\eta Z\langle 0|(\cdot)|0\rangle$, for $Z\equiv |0\rangle\langle 0|-|1\rangle\langle 1|$, and $|0\rangle$ and $|1\rangle$ are orthonormal states. Let $\cI$ denote the identity map. Then, for any $p\in[0,1]$, $\cE\equiv (1-p)\cI+p\cD$ and $\cE'\equiv(1-p)\cI+p\cD'$ saturate the inequality Eq.~\eqref{eq:app3}, up to a constant factor of 2: $|\Fmin'-\Fmin|= d\Vert\rho_{\cE'}-\rho_\cE\Vert_\mathrm{tr}$, for these two maps. Small $p$ values give channels that describe weak noise, the case of interest in this work.

\section{Estimation of classical computational resource cost}\label{sec:app2}
In each iteration of the GFM scheme, the classical computational resource is
spent on computing fidelity values with the DTFE scheme, followed by the
proposal of the next set of states with the CMA-ES algorithm. To estimate a single value of fidelity using the DTFE scheme, one would compute the average of $X$ over $d^2$ bases [see Eq.~\eqref{eq:fid2}]. For each basis, to estimate the value of $X$ from experimental data, one would take an average over the number of different outcomes, which is bounded by $d$. Therefore, the computational complexity of the DTFE scheme is $O(d^3)$. According to Ref.~\cite{CMAES4}, the computational complexity of each iteration of the CMA-ES algorithm is $O(d^2)$. Since the required number of function evaluations is observed empirically to be $O(d)$ to $O(d^2)$ \cite{CMAES3}, the classical computational resource cost of the GFM scheme is estimated to be $O(d^5)$.

The QPT fidelity estimation scheme is composed of three computational tasks, i.e., linear inversion, projection of the solution from linear inversion to the nearest CPTP map, and numerical minimization of fidelity of the reconstructed channel. In linear inversion, one tries to invert a system of linear equations, $Ax=f$. Here, $x$ is a $d^4 \times 1$ column vector of the parameters of the channel. $A$ is a $m \times d^4$ matrix determined by the input states and measurements, where $m$ denotes the total number of outcomes. For the product tetrahedron input states and measurements used in our scheme, $m=d^4$. $f$ is a $m \times 1$ column vector of the relative frequencies of the $m$ outcomes. Due to the use of product input states and product measurements in our scheme, the matrix $A$ can be written as a tensor product of matrices, i.e., $A = A_1 \otimes A_2 \otimes ... \otimes A_{2n}$. As a result, the computational cost is greatly reduced. Inverting the $2n$ matrices, $\{A_i\}_{i=1}^{2n}$ costs $O(n)$. The computation of $A^{-1}f = A_1^{-1} \otimes A_2^{-1} \otimes ... \otimes A_{2n}^{-1} f $ costs $O(nd^4)$ (see Appendix E of Ref.~\cite{superfastMLE}). 

The projection algorithm in Ref.~\cite{CPTPalgorithm} works by projecting the channel onto CP space and TP space alternatively and iteratively. In each iteration, the most computationally expensive operation is the eigendecomposition of the Choi state, which has the computational complexity of $O(d^6)$. It is, however, not clear how the number of iterations in the projection algorithm scales with the dimension. The computational complexity of the projection algorithm is thus estimated to be at least $O(d^6)$.

\makeatletter\global\advance\@colroom-10pt\relax\set@vsize\makeatother

The numerical minimization of the fidelity of the reconstructed channel is
performed using the conjugate-gradient algorithm. The calculation of the gradient in each iteration of the conjugate-gradient algorithm involves a matrix-vector multiplication which cost $O(d^4)$. The observed scaling of the number of iterations required for the conjugate-gradient algorithm implemented here is, at most, $O(d)$. Thus, the computational complexity of the numerical minimization is $O(d^5)$. Since the projection algorithm is the most expensive task among the three computational tasks, the classical computational resource cost of the QPT fidelity estimation scheme is estimated to be at least $O(d^6)$.




\begin{thebibliography}{99}
\bibitem{FN}
For the Class-HS example in Sec.~\ref{Examples}, we observe that the number of repeats we need for one- to five-qubit systems is 4, 7, 3, 1, and 1, respectively. For the Class-PA example in Sec.~\ref{Examples}, we observe that the number of repeats we need is 1 for one- to five-qubit systems.
\end{thebibliography}
\end{document}